# Industrial Edge-based Cyber-Physical Systems
## - Application Needs and Concerns for Realization

Martin Törngren[†], Haydn Thompson[††], Erik Herzog[†††], Rafia Inam[††††], James Gross and György Dán[†]

[†] KTH Royal Institute of Technology, Stockholm, Sweden (*martint, jamesgr, gyuri@kth.se*)
[††] Thinkk BV, Schiphol, Netherlands (*haydn.thompson@thhink.com*)
[†††] Saab Aeronautics, Linköping, Sweden (*erik.herzog@saabgroup.com*)
[††††] Ericsson, Stockholm, Sweden (*rafia.inam@ericsson.com*)

## ABSTRACT

Industry is moving towards advanced Cyber-Physical Systems (CPS), with trends in smartness, automation, connectivity and collaboration. We examine the drivers and requirements for the use of edge computing in critical industrial applications. Our purpose is to provide a better understanding of industrial needs and to initiate a discussion on what role edge computing could take, complementing current industrial and embedded systems, and the cloud. Four domains are chosen for analysis with representative use-cases; manufacturing, transportation, the energy sector and networked applications in the defense domain. We further discuss challenges, open issues and suggested directions that are needed to pave the way for the use of edge computing in industrial CPS.

## 1 Introduction

Edge computing has been described as one of the next logical steps in the ongoing digital transformation, towards a computing continuum. In addressing limitations of the cloud, and needs for locally available high performance computing, edge-computing systems are expected to have a tremendous application and projected market potential, see e.g. [1-5].

Edge computing has to our understanding so far mainly been driven by IT and OTT (Over-The-Top – cloud and media providers) as well as the Telecom sectors. Business opportunities in particular in content delivery networks (e.g. streaming media, gaming, web) drive these developments where edge computing in the form of distributed (localized) cloud data centers promises to provide technical benefits in terms of reduced latency, reduced transfer of data and lower energy usage, and improved privacy [1, 3]. The same opportunities with edge computing have also been highlighted in many other domains including smart manufacturing or Industry 4.0, smart cities, and transportation, see e.g. [15-18, 33].

Likely because of the promises in multiple application domains, and the potential to use different technologies for realizing edge computing, multiple visions of edge computing have been proposed and are being driven. These include *MEC - Multi-access Edge Computing* (strongly connected to telecommunications and 5G) [5], *fog computing* (relating to augmenting local computations of edge devices by exploiting communication devices such as routers and gateways in collaboration with the cloud) [4], and finally by means of *cloudlets* (small scale localized data centers) [6].

In this paper we are interested in industrial Cyber-Physical Systems (CPS) and what role edge computing could take in such systems, complementing current industrial and embedded systems, and the cloud, as CPS expand to become smarter, more automated, connected and collaborating [14].

While a number of studies of industrial requirements can be found in the literature, they often focus on specific domains or properties, with an emphasis on manufacturing in the context of fog computing, e.g. [2, 33-37]. Survey papers on edge computing in industrial applications reveal that the main body of research only to a limited extent considers key requirements of aspects of trustworthiness, [1-9, 33-37]. For example, [2] states, "most of the articles in the literature about fog computing do not consider failure or fault in the fog network.", [34] puts forward "Low-cost fault-tolerance and security … as open challenges", and [35] concludes, "aspects such as safety and security, …and their important interplay, have not been investigated in depth.".

A corresponding first goal of this paper is to elicit needs for industrial CPS across industrial domains, where the edge could play a role in realizing some of those needs; this is the topic of Section 2. The industrial needs and use cases largely stem from discussions brought forward in the TECoSA competence center and are further underpinned by various state of the art investigations, e.g. [19-21].

Edge computing still represents a relatively new concept, with strong potential in multiple domains and relevance for many stakeholders. There is uncertainty concerning what form future edge computing ecosystems will take, and which architectural approaches and standards will dominate. A second goal of this paper is therefore to identify and discuss challenges and directions for the use of edge computing in industrial CPS; we treat this topic in Section 3, and finally provide concluding remarks in Section 4.

## 2 Industrial CPS – Future Applications, Use Cases, Capabilities and Requirements

Among many potential industrial domains, we have chosen to investigate select but representative applications, part of the manufacturing, transportation, energy and defense sectors. Though substantially different at first sight, these applications share the same high-level requirements and challenges when it comes to their integration with edge computing.

Industry is facing a multifaceted transformation driven by sustainability needs and the ongoing digitalization. The former is, for example, manifested by increasingly stringent regulations regarding $CO_2$ emissions and producer responsibility, but also with end-user demands for sustainable products and services. The latter, digitalization, is characterized by trends towards increasingly connected, collaborating, automated and smart products and services. Importantly, digitalization enables to drive the transition to a circular economy by providing identification, monitoring, prediction, and tailored maintenance capabilities relying on CPS, [14]. The required transformations encompass entire supply chains and business models on the path towards circular systems.

In the following we discuss the selected industrial domains and use cases, and summarize common CPS application requirements.

### 2.1 Road Transportation

The automotive sector is seeing a radical shift where the role of electronics and software is growing massively driven by new transportation demands, business models, safety, and electrification. This change is creating a shift in vehicle models and shorter innovation cycles. New trends such as electromobility, automated driving, and modern mobility services require new technologies and have sociotechnical impact. The softwarization of vehicles, security threats, and acting in complex environments also drive frequent updates and upgrades. Challenges arise with more functions and services implemented in software, with increasing complexity of electronic component interoperability as well as with the management of agile and vulnerable supply chains in terms of logistics and production processes. To support this, integration platforms are needed to reduce development time and increase product quality w.r.t. traceability, safety, and cybersecurity. As a further source of complexity, there are different levels of real-time requirements for example for vehicle controls vs. for infotainment.

The long and winding road towards automated driving demonstrates the needs and opportunities for connectivity and collaboration, [22]. Data gathering through vehicles and a smart infrastructure will enable further innovation and requires the coordinated management (and regulation) of huge amounts of data. There are also opportunities for data sharing across sectors such as energy, road maintenance, mobility, well-being and renewables.

Fig. 1 highlights a potential future vision of an Intelligent Transport System (ITS), involving collaborating vehicles, a smart infrastructure and other road users. The vision indicates the potential role of various computing and communication technologies. Providing a smart infrastructure with perception capabilities and data sharing orchestrating of information among road users, promises to drastically improve road safety, traffic performance (in terms of reduced traveling times), minimizing energy usage, as well as using the data for proactive maintenance of the roads. Realizing the vision will require new capabilities such as new perception assets, additional computation, communication and "coordination", to be established in the road transport system.

Future vehicular communication systems (e.g. V2X and V2I), will need to act with distributed intelligence, as local computing power gives greater flexibility resulting in less data being sent to the cloud. Connectivity will be a core element of an integration platform.. As cars become more connected and autonomous, the data collected and transferred will rise rapidly leading to the need for local high performance and real-time computing. This requires management across multiple actors and at a global scale with filtering of data. Standards are needed for real-time, secure and safety-critical communication as part of the system design.

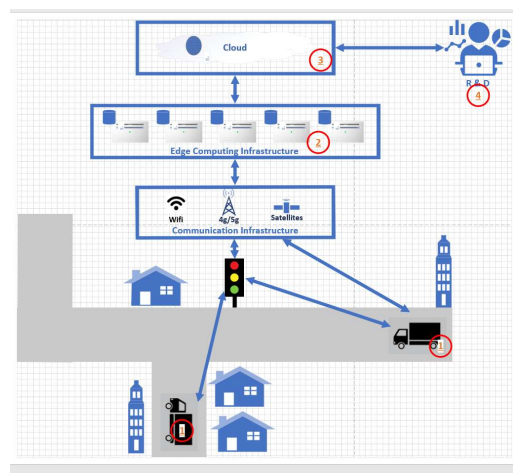

**Figure 1: Illustration of parts of an ITS; computational nodes are shown with markers: (1) automated vehicles, (2) local infrastructure, (3) cloud, and (4) developer/operator/ maintainer center. Figure courtesy of Naveen Mohan, KTH.**

### 2.2 Manufacturing

In manufacturing, optimizing production is key, and here computing, communication and coordination play important roles. Major industry issues are vendor lock-in and interfacing to legacy systems [10]. The use of open systems and standard technologies to replace proprietary systems would allow communication and control between machines from different vendors and the migration of legacy systems into new systems. There is a blurring of boundaries between the automation and enterprise worlds blending "Operational Technology" (OT) with Information Technology (IT) [11]. OT is often used in this context to refer to existing (dedicated) computer control and embedded systems, characterized by strong demands on robustness, real time and high availability, whereas IT is associated with data gathering, performance, security and enterprise system integration. Availability and cost-efficiency are strong drivers for predictive and preventive maintenance.

Although many of the technology building blocks exist, there is reluctance to put these on the shop floor. Some of the barriers could be overcome by establishing international standards for connectivity and communication, in particular, for data access and exchange [12]. However, there are also needs for computational performance beyond what is currently provided. A prime example of that is the execution of heavy AI-based algorithms and the processing a large amounts of data in the form of videos or images. Further, robots are usually resource constrained and thus need to be complemented with edge/cloud resources to meet the real-time requirements (on both communication and computation) [23].

As an example use case, consider a future smart manufacturing system, going from an industry where the robot is locked into a cell to a collaborating system where the robot is collaborating on the floor with human colleagues and other devices. Beyond collaboration, such a system would be characterized by ease of re-configurability, human augmentation and support through AR, a multitude of sensors to improve context awareness and predictive capabilities, and finally, interoperability with other parts of the manufacturing system. Such features, promise to enhance quality, safety and efficiency of the manufacturing system, as well as on-demand reconfigurability. Similar to the ITS use case, such a future manufacturing cell would also require new capabilities including perception, computation, communication and "coordination".

The safety needs for human-robot collaborative use cases are high and the robots/machines should respond and interact with humans in a timely manner. This places requirements on real-time communication and a certain network quality for the regular data exchange. Needs for local computations include perceiving the environment and extracting semantic-information to create a knowledge representation of the environment [24], and for understanding humans' behaviour including movement, direction and "mode" (e.g. distraction). This calls for novel architectures which can combine and orchestrate a number of devices, sensors and computational resources, in dynamic environments.

Potential further opportunities with edge computing include the ability to reduce cost and consolidate many different pieces of hardware condensing them on to server platforms at the local edge. Key needs are seamless access to sensor data in distributed systems, virtualization of multiple applications onto one device to allow control and safety functions to run together, and real time communication for the shop floor. Cloud-to-edge computing is becoming pervasive in areas such as preventive maintenance.

## 2.3 Energy

The energy sector is driven technically by latency, resiliency, privacy, regulation, and security. The uptake of renewables to generate power locally calls for a new energy system that needs an integrated approach over carriers and new governance creating a "perfect storm" in the marketplace [13]. The consequence is a move away from centralized grids (and separated producers vs. consumers), to a "prosumer" culture of Micro grids with local generation and energy storage capabilities that can either operate synchronously with the grid or act autonomously in island mode, adding a potential for improved resilience.

An additional major driver is the move towards electric cars, and the need for creating a new infrastructure for EV charging. This opens the opportunity to use electric car batteries as an energy storage source for local grids and also to provide power supply at peak demands for power balance. This trend coupled with the move to renewables is a game changer potentially transforming the electrical infrastructure rapidly as well as driving the need for cross-domain interaction between energy, automotive and building automation in future smart cities.

Looking forward there is thus also a need to cooperate and share data with the automotive and building sectors. This goes beyond just technical projects, requiring change management, stimulation of public private cooperation and demonstration of cross-silo collaboration at national/local levels.

Heating and cooling of buildings is a key consumer of energy and source of CO2 emissions. The increasing use of smart building automation, both for commercial premises and in the home, is driving a proliferation of sensors and intelligence being installed in buildings to more efficiently operate HVAC [2] systems and in household devices to reduce energy consumption and CO2 emissions. There is a need for interoperability standards to interconnect devices and for orchestration mechanisms to provide cross-domain control and optimization of energy.

There is a significant opportunity if the next generation smart meters are based on IoT and edge computing architectures. Privacy and regulatory concerns are also likely to keep enterprises from pushing their data or customer data to the cloud. For integration there is a need to move away from working in silos coupled with pull to adopt open solutions based on standards. It is also essential to understand the interfaces and what information needs to be published between different domains.

## 2.4 Defense

In the defense domain, the driving concern is speeding up the OODA (Observe, Orient, Decide, Act) decision-making loop to ensure a superior resolution and velocity compared to any potential adversary. To this end, a layered operational architecture where individual components can act both as device and data center nodes has emerged. For instance, an Airborne Early Warning (AEW) System has its own set of sensors, but also ample computational resources to fuse information from other device nodes, such as manned or unmanned fighter aircraft. At the same time, the AEW has the capability to relay information to ground based command and control centers. Consequently, rather than having distinct edge nodes there is a continuum – from devices over device-edge hybrids to data centers. As has historically been often the case, the defense sector anticipates already an infrastructure yet to emerge in the private sector. Real-time operation (speed and predictability), and perception capabilities (including sensor resolution) are key extra-functional characteristics in these applications. In the future it is expected that there will be a large increase in the number of devices connected, prompting the need for even more computational power. There is also a need to allow devices to connect dynamically raising the importance of trustworthiness – malicious devices must be detected and disconnected promptly. A further trend is that of directed communication to decrease the risk of detection – leading to more stringent real-time requirements – and of course the need for more computational resources.

---

[2] Heating, Ventilation and Air Conditioning

## 2.5 Application Commonalities

The domains just described are large fields in their own right, where specific requirements will be defined for sub-domains and applications. At a high level these domains however share many requirements and challenges. This is something we repeatedly notice: Despite what appears as quite different domains, our industrial stakeholders find it straightforward to identify similar challenges, often leading to research to address these. Based on an assessment of the described application domains, we summarize common categories of CPS application requirements in Table 1.

| Functionality, configuration & real-time | Dependability & trustworthiness |
|---|---|
| **(I) Context awareness and human-machine interaction:** Sensing needed for environment perception, allowing adaptation of CPS behaviour including communications, compute and algorithms to the instantaneous context. Human-machine interaction poses further requirements on awareness. | **(V) Safety:** Adequately reducing risk for harm to humans and the environment. An increasing complexity poses a challenge in introducing new risks and uncertainty. While new edge based solutions can contribute to enhance performance and safety, they could backfire and must be shown not to increase risks. |
| **(II) Algorithms:** Algorithms beyond legacy industrial control, including anomaly detection, augmented reality overlays, coordination planning, etc. | **(VI) Availability:** Prevent unplanned downtime due to failures and attacks by, e.g., predictive maintenance, resilient architectures and incident response. |
| **(III) Interoperability, adaptivity and scalability:** Includes needs to deal with technologies that can scale, handling multiple communication protocols, intermittent connectivity, mobility and failures, requiring adaptability for example through reconfigurations. | **(VII) Security:** Attacks must be carefully addressed and risks mitigated by design, intrusion detection and handling, and continuous life-cycle risk management. Ensuring both confidentiality (critical industrial data) as well as privacy (human users, operators, etc.) is essential. |
| **(IV) Real-time:** Closed-loop supervision and control systems need predictability, responsiveness and often also synchronized actions and correlation between multiple data streams/vents (requiring some notion of global time). | **(VIII) Transparency:** CPS need to embed notions of risk and strategies for dealing with them. Verification and failures further require transparency and explainability, e.g. through monitoring and means to understand the reasons for CPS behaviours. |

**Table 1. Industrial CPS – requirement categories**

The trend of increased capabilities of CPS to act (autonomously or partly autonomously) in more unstructured environments drives a need for improved context awareness, which in turn requires improved (and more) sensors, communications and storage. The availability of more data and information enables prediction (of for example intent of machine operators on the factory floor working together with robots) and planning for improved performance and for mitigating risks. The deployment of CPS in more complex environments further prompts interactions and integrations with more systems, including dynamic temporary connections, emphasizing the need for interoperability.

Cyber-physical systems act in the physical world and have to obey real-time constraints. Failing to meet these constraints may imply failures of functionality, which can result in loss of service and hazardous situations. Timing requirements go much beyond deadlines, relate to age of data, and timing has to consider the entire CPS. Timing requirements encompass both best effort and predictable/deterministic hard real-time systems, the latter in which a system is designed to provide predictable behavior. The actual real-time speed needed is naturally application dependent and could range from ms or less, in, for example, applications in telesurgery, and vehicle platooning, to seconds or more in control applications with slower closed loop dynamics.

The second column focuses on dependability and trustworthiness [20, 26-28]. Both these concepts can be seen to represent umbrella terms, embracing multiple properties. Dependability has traditionally been associated with aspects like reliability, availability, maintainability, safety and security. Trustworthiness has traditionally been associated with human-machine interactions and security – referring to how we as humans perceive trust in relation to services and machines. Trustworthiness is more recently evolving to become a broader umbrella term, subsuming dependability and adding AI inspired properties such as transparency, explainability and fairness [29]. In this context, the concept of assurance and liability become very important [30]; we elaborate further on safety/assurance cases in Section 3.3.

Current safety practices and standards are very stringent and costly for highly critical systems. This has led to a tradition in which a system is developed, verified, validated and certified, and then touched as little as possible after deployment. Moreover, it is difficult to use COTS software technologies, such as e.g. middleware, protocols, etc. that was not developed with safety in mind, [36]. This provides a huge challenge for future CPS which will require adaptation over time including upgrades to accommodate for learnings including new risks. It is not clear how this could be achieved in a cost-efficient way. It should be noted that availability is given its own entry in Table 1 due to its importance in industrial applications [40]. Availability is often also stated as a sub-attribute of security, referring to when availability is compromised, e.g., by denial of service attacks, [30].

It is also worthwhile to mention confidentiality, which is an often overlooked aspect of security in industrial control systems. Data related to operations could in practice be sensitive, as it can reveal production volumes, production quality, and thereby the profitability of a company. Data confidentiality can in fact represent an obstacle to equipment vendors for providing cloud-based data-driven services for, e.g., predictive maintenance, quality assurance, etc. Data confidentiality is often mentioned as a key driver for adopting federated learning for data driven services, but due to recent research results on property inference attacks against machine learning models [46], the industry acceptance of federated learning is slower than one may have expected.

The categories of requirements in Table 1 are described on a high level. More detailed and quantitative requirements will be related to specific applications and contexts. For example, regarding safety and security, these will depend on the level of risk and criticality involved. Safety standards such as IEC61508 adopt a typical risk classification scheme, associating systems, functions and components with particular integrity levels that are accompanied with corresponding requirements (process, technical, organizational) and reliability/availability targets. For example, for the highest safety integrity level in IEC61508, the required probability of failure/h is $10^{-8} – 10^{-9}$, [41].

We argue that many of the requirements in Table 1 will benefit from, and in some cases only be feasible with, various realizations of edge computing. In other words, for future CPS to become more capable, autonomous and collaborating, the needs for localized high performance computing will be increasing. This will not remove the need for smaller embedded systems nor the cloud, but will add a complementary layer. As indicated in Table 1, however, this will require specific attention to new safety and security risks associated with new edge-based CPS including addressing new types of failure modes and attack surfaces.

## 3 Challenges, Open Issues and Directions in Adopting Edge Computing for Industrial CPS

In addressing the second goal of the paper we here discuss opportunities, challenges, open issues and directions needed to pave the way for the use of edge computing in industrial CPS.

### 3.1 Taxonomy Need and Overcoming Community Differences

Given the relative novelty of edge computing it is not surprising that several interpretations have developed as to what edge computing means, also considering that very different stakeholders and communities are involved. However, this also causes some confusion that complicates the dialogue, as exemplified by very distinct interpretations and the multitude of terms representing variants of edge computing (near edge, far edge, nano edge, enterprise edge, multi-access edge, cloudlets, fog computing, to name a few). Moreover, in the current discourse, edge computing can be associated with either locality, computing technologies, or both, [1]. To discuss and investigate what role edge computing could play for future industrial CPS, we believe that further interactions between disparate communities is essential. We also think that an effort to try to create a taxonomy and a common terminology would be relevant to undertake, albeit challenging! Several of the early interpretations of edge computing are based on the idea to leverage advances in cloud computing by providing small-scale cloud data centers, located at the network edge (internet) [3]. The concept of edge has also come to be used in another flavor in embedded and industrial systems domains, referring to (the expansion of) computational capabilities within embedded systems; seen as the "device edge". Such solutions still differ widely from small-scale cloud data centers.

As we embark towards a computing continuum, we believe that the role of the various types of computing systems and their interactions need to be embraced, including,

- embedded systems, from traditional dedicated functions and resource constrained systems to embedded data centers, as emerging in highly automated vehicles.
- edge data centers, representing localized centers whether as new stand-alone deployments or as part of/integrated with the 5G network, and,
- cloud data centers (centralized and distributed).

In industrial systems, this computing continuum will also integrate sensors, actuators, human-machine interfaces, etc. beyond computing and communication, providing a variety of capabilities, characterized by different geographical locations, and differences in dynamicity and mobility of the involved actors/systems.

We believe that Table 1 represents a useful set of requirement categories that can serve as a starting point for a taxonomy together with system and application characteristics such as dynamicity, mobility and locality.

### 3.2 Edge Computing Opportunities and Innovation Eco-systems

The combination of advances in various technologies related to CPS provide unprecedented opportunities for innovation and new services, based on for example improved perception, awareness, prediction, planning and other analytics capabilities. Innovation can also be accomplished through new business models, often associated with a service orientation. For future CPS there is thus a potential for both incremental and disruptive innovations.

To pave the way for such new edge-based CPS, there are needs to create long-term knowledge and innovation eco-systems, representing collaborations between the (evolving set of) stakeholders that are critical for innovation, [43]. We know that large-scale technological developments take time, [25], and the underlying pattern of that is true also for edge-based CPS where multiple types of knowledge, educated/trained people, components, systems, business models, legislation, companies and collaborations will be needed. In many cases the new types of systems can be characterized as System of Systems (SoS), in which there is no single system integrator and where the constituent units (cmp. e.g. road infrastructure, cars, communication networks in an ITS) evolve independently, [44].

Some important considerations for establishing such new innovation eco-systems include

- establishing collaborative testbeds and data sharing, as important for gaining new knowledge in going beyond existing engineering methodologies, [14, 42],
- considering business models together with technological solutions for edge computing, e.g. regarding quality of service, resource sharing and corresponding contracts,
- legal considerations including how to deal with data confidentiality/privacy, assurance and liability.

There are multiple interesting directions regarding technological innovations. The presented prospects for edge computing, including e.g. MEC [5], cloudlets [3] and fog computing [4], point to different potential technological solutions. A key difference between cloud computing and edge computing is that in cloud computing the services tend to be general purpose and application agnostic. However, when working at the edge it is important to understand the functional and extra-functional requirements for the application. Here the winners in new emerging markets are likely to be the ones who understand the sectorial requirements of key industrial sectors such as automotive, energy, manufacturing, etc. This needs to build upon expertise and advances in embedded sensors and system design, computing, networks, 5G, microprocessors and artificial intelligence.

A central starting point for establishing edge computing in industrial applications is to establish collaborations. There are several examples of such initiatives, including the Automotive Edge Computing Consortium (AECC) [31], which gathers stakeholders to drive the evolution of edge network architectures and computing infrastructures to support high volume data services for future connected vehicles. Another example is the TECoSA competence center, which gathers industrial partners from multiple domains, with a particular emphasis on trustworthiness of future edge computing systems and applications [32].

### 3.3 Trustworthiness and Dependability

Introducing edge computing into industrial CPS necessitates a strong focus on trustworthiness, requiring that critical services maintain availability, avoid hazardous failures, and do not violate essential agreements and legislations. Many industrial CPS are subject to approvals and certification (e.g. type approval/homologation for cars) and thus have to "demonstrate" trustworthiness upfront. In safety-critical CPS, it is already a fact that so called safety cases have to be developed, e.g. as part of certifying systems before they are released for use (see, e.g. [30]). A safety case (as a safety-related instance of a more general assurance case that could also refer to other properties such as security) should provide "a structured argument, supported by a body of evidence, that provides a compelling, comprehensible and valid case that a system is safe …" (quote from NASA System Safety Handbook ver. 1, 2014). To be efficient, safety case activities have to be integrated with other life-cycle processes.

Augmenting CPS through more advanced computing and communications will enhance their capabilities and also their complexity. This complexity needs to be appropriately managed, and will require stringent efforts throughout the life cycle, including development (designing the proper architecture and mechanisms), operation (monitoring and maintenance, ensuring proper organizational roles), and finally failure management including attention to reporting and forensics.

In embarking towards increasingly capable and complex edge-based CPS, we believe that trustworthiness needs to be incorporated as a first-class citizen in research and endeavors focused on industrial edge-based CPS. Such efforts need to consider the multiple interdependent attributes of trustworthiness including new methodologies and architectures to address them.

## 4 Conclusions

The potential for providing CPS with entirely new capabilities provides opportunities for edge computing in many key industrial sectors. Edge computing and networking is also creating enablers for connections between sectors to create new integrated services driven by ITS, energy, etc. to provide a more efficient society while addressing climate change goals.

Many challenges still remain, and beyond technological ones, include contractual, privacy, security, liability, safety assurance, and corresponding standards. There is a need to stimulate community interactions in the first place, and to further promote collaborations to explore and evaluate future edge-based industrial CPS including through testbeds. Finally, in introducing edge computing in industrial CPSs, it is key that trustworthiness is treated as a priority and first-class citizen.

### ACKNOWLEDGMENTS

This work has been carried out as part of the Vinnova Competence Center for Trustworthy Edge Computing Systems and Applications at KTH Royal Institute of Technology. We acknowledge useful insights contributed by the TECoSA partners, especially from Atlas Copco (Martin Karlsson, Lifei Tang and Sofia Olsson), Diarmuid Corcoran from Ericsson, and Naveen Mohan from KTH.